\def\be{\begin{equation}}
\def\ee{\end{equation}}
\def\ba{\begin{array}}
\def\ea{\end{array}}
\def\t{\tilde}
\def\p{\prime}
\def\Rb{{I\!\! R}}

\documentstyle[12pt]{article}
\begin{document}
\parskip=3pt
\parindent=18pt
\baselineskip=20pt
\setcounter{page}{1}
\centerline{\Large\bf A Remark on Integrable Poisson Algebras}
\bigskip
\centerline{\Large\bf and Two Dimensional Manifolds}
\vspace{6ex}
\centerline{\large{\sf Sergio Albeverio$^\star$} ~~~and~~~ {\sf Shao-Ming Fei}
\footnote{\sf Alexander von Humboldt-Stiftung fellow.\\
\hspace{5mm}On leave from Institute of Physics, Chinese Academy of Sciences,
Beijing}}
\vspace{4ex}
\parindent=40pt
{\sf Institute of Mathematics, Ruhr-University Bochum,
D-44780 Bochum, Germany}\par
\parindent=35pt
{\sf $^\star$SFB 237 (Essen-Bochum-D\"usseldorf); 
BiBoS (Bielefeld-Bochum);\par
\parindent=40pt
CERFIM Locarno (Switzerland)}\par
\vspace{6.5ex}
\parindent=18pt
\parskip=5pt
\begin{center}
\begin{minipage}{5in}
\vspace{3ex}
\centerline{\large Abstract}
\vspace{4ex}
The relations between integrable Poisson algebras with three 
generators and two-dimensional manifolds are investigated. 
Poisson algebraic maps are also discussed.
\end{minipage}
\end{center}

\newpage
Poisson algebras have been discussed widely in Hamiltonian mechanics and
in the quantization of classical systems such as canonical quantization
and Moyal product quantization, see e.g. \cite{found,moyal,poisson,poisson1}.
In this paper we investigate the relations between manifolds and Poisson
algebras. We find that there exist general
relations between integrable Poisson algebras with three 
generators and two-dimensional smooth manifolds, which gives manifest
geometric meanings to the Poisson algebras and from which
the Poisson algebraic maps can easily be discussed.

A symplectic manifold $(M,\omega)$ is an even 
dimensional manifold $M$ equipped with a symplectic two form $\omega$, 
see e.g. \cite{geom,wood,hofer,symp}. Let $d$ denote the exterior derivative on $M$. 
By definition a symplectic form $\omega$ on $M$ is closed, 
$d\omega=0$, and non-degenerate, $X\rfloor\omega=0~~\Rightarrow~~ X=0$,
where $X$ is a (smooth) vector on $M$ and $\rfloor$ denotes the left inner
product defined by $(X\rfloor\omega)(Y)=\omega(X,Y)$
for any two vectors $X$ and $Y$ on $M$. The non-degenerancy means that
for every tangent space $T_xM$, $x\in M$, $X\in T_xM$, if $\omega_x(X,Y)=0$
for all $Y\in T_xM$, then $X=0$.

Infinitesimal symplectic diffeomorphisms are given by vectors.
A vector $X$ on $M$ corresponds to an infinitesimal
canonical transformation if and only if the Lie derivative of 
$\omega$ with respect to $X$ vanishes,
\be\label{a1}
{\cal L}_{X}\omega=X\rfloor d\,\omega+d\,(X\rfloor\omega)=0\,.
\ee
A vector $X$ satisfying (\ref{a1}) is said to be a Hamiltonian vector field.

Since $\omega$ is closed, it follows from (\ref{a1}) that a vector $X$ is
a Hamiltonian vector field if and only if $X\rfloor\omega$ is closed.
Since $\omega$ is non-degenerate, this gives rise to an isomorphism between
vector fields $X$ and one forms on $M$ given by $X\to X\rfloor\omega$.
Let ${\cal F}(M)$ denote the real-valued smooth functions on $M$. 
For an $f\in {\cal F}(M)$, there exists a Hamiltonian vector field $X_{f}$
(unique up to a sign on the right hand of the following equation) satisfying
\begin{equation}\label{3}
X_{f}\rfloor\omega=-d\,f\,.
\end{equation}
We call $X_f$ the Hamiltonian vector field associated with $f$.

Let $f,g\in {\cal F}(M)$, $X_f$ and $X_g$ be the Hamiltonian vector fields
associated with $f$ and $g$ respectively. The Lie bracket $[X_f\,,X_g]$ 
is the Hamiltonian vector field of $\omega(X_f,X_g)$, in the sense that
$$
\ba{rcl}
[X_f\,,X_g]\rfloor\omega
&=&{\cal L}_{X_f}(X_g\rfloor\omega)-X_g\rfloor({\cal L}_{X_f}\omega)\\[4mm]
&=&{X_f}\rfloor d(X_g \rfloor\omega)+d(X_f\rfloor(X_g\rfloor\omega))
-X_g\rfloor d(X_f\rfloor\omega)\\[4mm]
&=&-d(\omega(X_f,X_g))\,,
\ea
$$
where the Cartan's formula for the Lie derivative ${\cal
L}_{X}=i_X\circ d+d\circ i_X$ of a vector $X$ has been used.
The function $-\omega(X_f,X_g)$ is called the Poisson bracket of $f$ and $g$
and denoted by $[f,g]_{P.B.}$,
\begin{equation}\label{4}
[f,g]_{P.B.}=-\omega(X_{f},X_{g})=-X_{f}\,g.
\end{equation}
Since $\omega$ is closed, the so defined 
Poisson bracket satisfies the Jacobi identity
$$
[f,[g,h]_{P.B.}]_{P.B.}+[g,[h,f]_{P.B.}]_{P.B.}+[h,[f,g]_{P.B.}]_{P.B.}=0\,.
$$
Therefore under the Poisson bracket operation the space $C^\infty(M)$ of
all smooth functions on $(M,\omega)$ is a Lie algebra, called the Poisson
algebra of $(M,\omega)$.

In general one calls Poisson algebra any associative, commutative algebra
${\cal A}$ over $\Rb$ with unit, equipped with a bilinear map
$[,]_{P.B.}$, called Poisson bracket satisfying
$$
\ba{ll}
1)~Antisymmetry&[f,g]_{P.B.}=-[g,f]_{P.B.},\\[3mm]
2)~Derivation~property&[fg,h]_{P.B.}=f[g,h]_{P.B.}+g[f,h]_{P.B.},\\[3mm]
3)~Jacobi~identity&[f,[g,h]_{P.B.}]_{P.B.}+[g,[h,f]_{P.B.}]_{P.B.}
+[h,[f,g]_{P.B.}]_{P.B.}=0,
\ea
$$
for any $f,g,h\in {\cal A}$. 

Now let ${\cal A}$ be a Poisson algebra with three generators
$(x_1,x_2,x_3)=x$ and a Poisson bracket of the form
\be\label{1}
[x_i,x_j]_{P.B.}=\sum_{k=1}^3\epsilon_{ijk}f_k\,,
\ee
where $\epsilon_{ijk}$ is the completely antisymmetric tensor and
$f_i$, $i=1,2,3$, are smooth real-valued functions of $x$,
restricted to satisfy the Jacobi identity:
$$
\ba{l}
[x_1,[x_2,x_3]_{P.B.}]_{P.B.}+[x_2,[x_3,x_1]_{P.B.}]_{P.B.}
+[x_3,[x_1,x_2]_{P.B.}]_{P.B.}\\[4mm]
=\displaystyle\frac{\partial f_1}{\partial x_2}f_3
-\displaystyle\frac{\partial f_1}{\partial x_3}f_2+
\displaystyle\frac{\partial f_2}{\partial x_3}f_1
-\displaystyle\frac{\partial f_2}{\partial x_1}f_3+
\displaystyle\frac{\partial f_3}{\partial x_1}f_2
-\displaystyle\frac{\partial f_3}{\partial x_2}f_1=0\,.
\ea
$$

{\sf [Definition 1].} The Poisson algebra (\ref{1}) is said to be integrable
if $f_i$ satisfies
\be\label{2}
\frac{\partial f_i}{\partial x_j}=\frac{\partial f_j}{\partial
x_i}\,,~~~ i,j=1,2,3.
\ee

Obviously the integrability condition (\ref{2}) is a sufficient condition for 
the Poisson algebra (\ref{1}) to satisfy the Jacobi indentity. 

Let ${\cal F}$ be the space of smooth real-valued functions of $x$,
$x\in \Rb^3$. We consider the realization of the Poisson algebra ${\cal A}$
in $\Rb^3$ and will not distinguish between 
the symbols $x_i$ of the coordinates of
$\Rb^3$ and the generators of ${\cal A}$. In the following $M$
will always denote a smooth two dimensional manifold 
smoothly embedded in $\Rb^3$.

{\sf [Theorem 1].} For a given integrable Poisson algebra ${\cal A}$
there exists a two dimensional symplectic manifold $M$ 
described by an equation of the form $F(x)=c$, $x\in \Rb^3$, 
with $F\in {\cal F}$ and $c$ an arbitrary real number, such that the
Poisson algebra generated by the coordinates of $M$ coincides with the
algebra ${\cal A}$.

{\sf [Proof].} A general integrable Poisson algebra is of the form
(\ref{1}),
$$
[x_i,x_j]_{P.B.}=\sum_{k=1}^3\epsilon_{ijk}f_k,
$$
where $f_i$, $i=1,2,3$, satisfy the integrability condition (\ref{2}).
What we have to show is that this Poisson algebra can be described by
the symplectic geometry on a suitable two dimensional symplectic 
manifold ($M,\omega$), in
the sense that the above Poisson bracket can be described by the
formula (\ref{4}), i.e., the Poisson bracket $[x_i,x_j]_{P.B.}$ is 
given by the Hamiltonian
vector field $X_{x_i}$ associated with $x_i$ such that
\be\label{th1}
[x_i,x_j]_{P.B.}=-X_{x_i}x_j=\sum_{k=1}^3\epsilon_{ijk}f_k,
\ee
with $x_i$ the coordinates of the two dimensional manifold $M$ in $\Rb^3$. 

Let $X_{x_i}^\p\in\Rb^3$ be given by
\be\label{6}
X_{x_i}^\p\equiv\sum_{j,k=1}^3\epsilon_{ijk}f_j\frac{\partial}{\partial x_k}.
\ee
Then $X_{x_i}^\p$ satisfies
$$
[x_i,x_j]_{P.B.}=-X_{x_i}^\p x_j=\sum_{k=1}^3\epsilon_{ijk}f_k.
$$

A general two form on $\Rb^3$ has the form
\be\label{5}
\omega^\p=-\frac{1}{2}\sum_{i,j,k=1}^3\epsilon_{ijk}h_idx_j\wedge dx_k\,,
\ee
where $h_i\in {\cal F}$, $i=1,2,3$.

We have to prove that $x$ can be restricted to a suitable 2-dimensional manifold 
$M\subset\Rb^3$ in such a way that $X_{x_i}^\p$ coincides with the
Hamiltonian vector field $X_{x_i}$ and $\omega^\p$ is the
corresponding symplectic form $\omega$ on $M$.

A two form on a two dimensional
manifold is always closed. What we should then check is that
there exists $M\subset\Rb^3$ such that for $x$ restricted to $M$
the formula (\ref{3}) holds for $f=x_i$, i.e.,
\be\label{a2}
X_{x_i}^\p\rfloor\omega^\p=-dx_i~~~~x_i\in M,~~~i=1,2,3.
\ee

Substituting formulae (\ref{5}) and (\ref{6}) into (\ref{a2}) we get
$$
\ba{rcl}
X_{x_i}^\p\rfloor\omega^\p&=&
-\displaystyle\sum_{j,k=1}^3\epsilon_{ijk}f_j\displaystyle
\frac{\partial}{\partial x_k}\rfloor
\displaystyle\frac{1}{2}\displaystyle\sum_{l,m,n=1}^3
\epsilon_{lmn}h_ldx_m\wedge dx_n\\[4mm]
&=&-\displaystyle\frac{1}{2}\displaystyle
\sum_{lmnjk}^3\epsilon_{ijk}\epsilon_{lmn}
f_jh_l(\delta_{km}dx_n-\delta_{kn}dx_m)\\[4mm]
&=&-\displaystyle\sum_{lnjk}^3\epsilon_{ijk}\epsilon_{lkn}f_jh_ldx_n
=-dx_i\,.
\ea
$$
That is,
\be\label{7}
\ba{l}
(1-f_2h_2-f_3h_3)dx_1+f_2h_1dx_2+f_3h_1dx_3=0,\\[3mm]
(1-f_3h_3-f_1h_1)dx_2+f_3h_2dx_3+f_1h_2dx_1=0,\\[3mm]
(1-f_1h_1-f_2h_2)dx_3+f_1h_3dx_1+f_2h_3dx_2=0.
\ea
\ee

Let us now look at the coefficient determinant $D$ of the $dx_i$ in
the system (\ref{7}). By a suitable choice of $h_1,h_2,h_3$ we can
obtain that $D$ is zero. This is in fact equivalent with the equation
\be\label{8}
f_1h_1+f_2h_2+f_3h_3=1
\ee
being satisfied. The fact that $D=0$ implies that there exists indeed an
$M$ as above. 

Substituting condition (\ref{8}) into (\ref{7}) we get
\be\label{9}
f_1dx_1+f_2dx_2+f_3dx_3=0\,.
\ee

From the assumption (\ref{2}) we know that the differential
equation (\ref{9}) is exactly solvable, in the sense that
there exists a smooth (potential) function $F\in {\cal F}$ and a
constant $c$ such that
\be\label{10}
F(x)=c
\ee
and $\partial F/\partial x_i=f_i$. The above manifold $M$ is then
described by (\ref{10}).

Therefore for any given integrable Poisson algebra ${\cal A}$ there 
always exists a two dimensional manifold of the form (\ref{10}) on which
$X_{x_i}^\p$ in (\ref{6}) is a Hamiltonian vector field and 
the Poisson bracket of the algebra ${\cal A}$ is given by $X_{x_i}^\p$ 
according to the formula (\ref{4}),
$[x_i,x_j]_{P.B.}=-X_{x_i}^\p x_j=\sum_{k=1}^3\epsilon_{ijk}f_k$.

The two dimensional manifold defined by (\ref{10}) is unique (once $c$ is
given). Hence an integrable
Poisson algebra is uniquely given by the two dimensional
manifold $M$ described by $F(x)=c$. $\rule{2mm}{2mm}$

Before going over to investigate the Poisson algebraic structures on general
two dimensional manifolds, we would like to make some remarks on special
symplectic properties of two dimensional manifolds.

{\sf [Proposition 1].} Let $M$ be a two dimensional manifold embedded in
$\Rb^3$. If $\omega$ is a symplectic form on $M$, then for
$\alpha(x)\neq 0$, $\forall x\in \Rb^3$, $\alpha^{-1}\omega$ is also a
symplectic form on $M$.

{\sf [Proof].} As $M$ is two dimensional, the two form $\alpha^{-1}\omega$
is closed, i.e., $d(\alpha^{-1}\omega)=0$, and $\alpha^{-1}\omega$ is 
nondegenerate, as $\omega$ is
nondegenerate. Hence $\alpha^{-1}\omega$ is also a symplectic form on
$M$. $\rule{2mm}{2mm}$

{\sf [Proposition 2].} For $f,g,h\in{\cal F}(M)$, if $[f,g]_{P.B.}=h$ on the
symplectic manifold $(M,\omega)$, then $[f,g]_{P.B.}=\alpha h$ on the
symplectic manifold $(M,\alpha^{-1}\omega)$, $\alpha(x)\neq 0$, $\forall
x\in \Rb^3$.

{\sf [Proof].} From formulae (\ref{3}) and (\ref{4}), on the
symplectic manifold $(M,\omega)$ the symplectic vector field $X_f$ satisfies 
$X_{f}\rfloor\omega=-d\,f$ and $[f,g]_{P.B.}=-X_{f}\,g=h$. If $\omega$
is changed to be $\alpha^{-1}\omega$, then $X_f$ becomes $\alpha X_f$ such
that $\alpha X_{f}\rfloor\alpha^{-1}\omega=-d\,f$. Therefore on 
$(M,\alpha^{-1}\omega)$, $[f,g]_{P.B.}=-\alpha X_{f}\,g=\alpha h$.
$\rule{2mm}{2mm}$

From the properties in Propositions 1 and 2 we define an equivalent class
of Poisson algebras on two dimensional manifolds.

{\sf [Definition 2].} On a two dimensional manifold embedded in
$\Rb^3$, a Poisson algebra $A$ is equivalent to a Poisson
algebra $B$ if the Poisson bracket on $A$ is the same as
the one on $B$, multiplied by some common non zero factor
$\alpha(x)$, $\forall x\in\Rb^3$.

{\sf [Theorem 2].} For a given smooth two dimensional manifold $M$ embedded 
in $\Rb^3$ of the form $F(x)=0$, $F\in{\cal F}$, $x\in \Rb^3$,
$x$ generates a Poisson algebra with the following Poisson bracket:
\be\label{t2}
[x_i,x_j]_{P.B.}=\sum_{k=1}^3\epsilon_{ijk}
\frac{\partial F(x)}{\partial x_k}.
\ee
This is unique in the sense of the equivalence in definition 2.

{\sf [Proof].} We asssume the symplectic form $\omega$ and the Hamiltonian 
vector field $X_x$ on $M$ are of the following general form:
$$
\omega=
-\frac{1}{2}\sum_{i,j,k=1}^3\epsilon_{ijk}h_i^{\prime}dx_j\wedge dx_k\,,
$$
\be\label{xi1}
X_{x_i}=\sum_{j,k=1}^3\epsilon_{ijk}f_j^{\prime}\frac{\partial}{\partial
x_k}\,,~~~~i=1,2,3,
\ee
$h_i^{\prime}, f_i^{\prime}\in {\cal F}$, $i=1,2,3$. 
From $X_{x_i}\rfloor\omega=-dx_i$ we have 
\be\label{11}
\ba{l}
(1-f^{\prime}_2h^{\prime}_2-f^{\prime}_3h^{\prime}_3)dx_1
+f^{\prime}_2h^{\prime}_1dx_2+f^{\prime}_3h^{\prime}_1dx_3=0,\\[3mm]
(1-f^{\prime}_3h^{\prime}_3-f^{\prime}_1h^{\prime}_1)dx_2
+f^{\prime}_3h^{\prime}_2dx_3+f^{\prime}_1h^{\prime}_2dx_1=0,\\[3mm]
(1-f^{\prime}_1h^{\prime}_1-f^{\prime}_2h^{\prime}_2)dx_3
+f^{\prime}_1h^{\prime}_3dx_1+f^{\prime}_2h^{\prime}_3dx_2=0,
\ea
\ee
where $dx$ are not independent since $F(x)=0$ implies
\be\label{12}
\sum_{i=1}^3\frac{\partial F(x)}{\partial x_i}dx_i=0\,.
\ee
Therefore the coefficient determinant of the system (\ref{11})
is zero, which gives
$$
\sum_{i=1}^3f^{\prime}_ih^{\prime}_i=1\,.
$$
Hence the system of equations (\ref{11}) becomes
\be\label{13}
\sum_{i=1}^3f^{\prime}_idx_i=0\,.
\ee

Equations (\ref{12}) and (\ref{13}) give rise to
\be\label{14}
f^{\prime}_i(x)=\alpha(x)\frac{\partial F(x)}{\partial x_i}\,,~~~~~i=1,2,3,
\ee
where $\alpha(x)\neq 0$, $\forall x\in\Rb^3$.

From (\ref{14}) the Hamiltonian vector field (\ref{xi1}) takes the form
\be\label{xi2}
X_{x_i}=\alpha(x)\sum_{j,k=1}^3\epsilon_{ijk}
\frac{\partial F(x)}{\partial x_j}\frac{\partial}{\partial x_k}\,.
\ee
Using formula (\ref{4}) we have
\be\label{15}
[x_i,x_j]_{P.B.}=\alpha(x)\sum_{k=1}^3\epsilon_{ijk}
\frac{\partial F(x)}{\partial x_k}\,.
\ee
This is just the formula (\ref{t2}) under the equivalence definition 2 of 
Poisson algebras on two dimensional embedded manifolds. $\rule{2mm}{2mm}$

It should be noted that if $F(x)=0$ defines a smooth two dimensional
manifold $M$ in $\Rb^3$, then $\alpha(x)F(x)=0$, $\alpha(x)\neq 0$,
$\forall x\in \Rb^3$, also defines the same manifold $M$. By formula
(\ref{t2}) we see that $F(x)=0$ and $\alpha(x)F(x)=0$ give rise to the same
Poisson algebra under the algebraic equivalence given by definition 2.

As $F\in{\cal F}$, we have that
$$
\frac{\partial}{\partial x_j}\left(\frac{\partial F(x)}{\partial x_i}\right)
=\frac{\partial}{\partial x_i}
\left(\frac{\partial F(x)}{\partial x_j}\right)\,,~~~~i,j=1,2,3.
$$
Therefore the Poisson algebra given by (\ref{t2}) is by definition integrable 
and it is uniquely given by the manifold $M$.

{\sf [Definition 3].}  $C\in {\cal F}$ is said to be a center of 
the Poisson algebra (\ref{1}) if it satisfies $[x_i,C]_{P.B.}=0$, $i=1,2,3$.

{\sf [Corollary 1].} The center elements of the Poisson algebra 
on the two dimensional manifold $F(x)=0$ are ({\bf 1}, $F(x)$).

{\sf [Proof].} From formulae (\ref{4}) and (\ref{xi2}) we have
$$
[x_i,F(x)]_{P.B.}=-X_{x_i}F(x)=\sum_{j,k=1}^3\epsilon_{ijk}
\frac{\partial F(x)}{\partial x_j}\frac{\partial F(x)}{\partial x_k}=0\,,
~~~~i=1,2,3,
$$
from which the proof follows. $\rule{2mm}{2mm}$

{\sf [Corollary 2].} For $f,g\in {\cal F}$, if $x$ satisfies the Poisson
algebraic relations (\ref{t2}), then
\be\label{16}
[f,g]_{P.B.}(x)=-\sum_{i,j,k=1}^3\epsilon_{ijk}
\frac{\partial f}{\partial x_i}
\frac{\partial F(x)}{\partial x_j}
\frac{\partial g}{\partial x_k}\,.
\ee

{\sf [Proof].} The Hamiltonian vector field associated with $x_i$ 
giving rise to the algebra
(\ref{15}) is given by formula (\ref{xi2}) and satisfies
\be\label{heh1}
X_{x_i}\rfloor\omega=-dx_i\,.
\ee
On the other hand the 
Hamiltonian vector field $X_f$ associated with $f$ satisfies by definition
\be\label{heh2}
X_{f}\rfloor\omega=-df\,.
\ee
From (\ref{heh1}) and (\ref{xi2}) the right hand side of 
equation (\ref{heh2}) reads
$$
\ba{rcl}
-df&=&-\displaystyle\sum_{i=1}^3
\displaystyle\frac{\partial f}{\partial x_i}dx_i
=\displaystyle\sum_{i=1}^3\displaystyle
\frac{\partial f}{\partial x_i}(X_{x_i}\rfloor\omega)\\[4mm]
&=&\displaystyle\sum_{i=1}^3\displaystyle\frac{\partial f}{\partial x_i}
\left(\displaystyle\sum_{j,k=1}^3\epsilon_{ijk}
\displaystyle\frac{\partial F(x)}{\partial x_j}
\displaystyle\frac{\partial}{\partial x_k}
\rfloor\omega\right)\,.
\ea
$$
Substituting it into (\ref{heh2}) we have, as $\omega$ is non degenerate,
$$
X_{f}=\sum_{i,j,k=1}^3\epsilon_{ijk}\frac{\partial f}{\partial x_i}
\frac{\partial F}{\partial x_j}\frac{\partial}{\partial x_k}\,.
$$
Therefore by definition
$$
[f,g]_{P.B.}(x)=-X_f g=-\sum_{i,j,k=1}^3\epsilon_{ijk}
\frac{\partial f}{\partial x_i}
\frac{\partial F}{\partial x_j}
\frac{\partial g}{\partial x_k}\,.
$$
{\hfill $\rule{2mm}{2mm}$}

Theorem 1 and Theorem 2 establish relations between two
dimensional smooth manifolds and Poisson algebras with three
generators. In what follows we study some properties related to smooth 
Poisson algebraic maps.

Let $A$ resp. $B$ be two integrable Poisson algebras with related 
two dimensional
manifolds $M_A$ resp. $M_B$ defined by $F_A(x)=0$ resp. $F_B(y)=0$ in
$\Rb^3$, where $x=(x_1,x_2,x_3)$ resp. $y=(y_1,y_2,y_3)$ are the 
generators of the algebra $A$ resp. $B$.

{\sf [Therom 3].} If the smooth algebraic map
$\t{y}(x)=(\t{y}_1,\t{y}_2,\t{y}_3)(x)$ 
gives rise to the algebra $B$, then $\t{y}$ satisfies $F_B(\t{y})=0$.

{\sf [Proof].} From Theorem 2 the Poisson algebra $A$ is given by
$$
[x_i,x_j]_{P.B.}=\sum_{i,j,k=1}^3\epsilon_{ijk}
\frac{\partial F_A(x)}{\partial x_k}\,.
$$
Using formula (\ref{16}) of Corollary 2 we have
\be\label{17}
[\t{y}_i,\t{y}_j]_{P.B.}(x)=
-\sum_{l,m,n=1}^3\epsilon_{lmn}
\frac{\partial \t{y}_i}{\partial x_l}
\frac{\partial F_A}{\partial x_m}
\frac{\partial \t{y}_j}{\partial x_n}\,.
\ee
Since  $F_A(x)=0$, we have that the $x_i$, $i=1,2,3$, are not 
independent. Without loosing 
generality we take $x_1$ and $x_2$ to be independent. By using the
relation 
$$
\sum_{i=1}^3\frac{\partial F_A(x)}{\partial x_i}dx_i=0\,,
$$
equation (\ref{17}) becomes
\be\label{17p}
[\t{y}_i,\t{y}_j]_{P.B.}(x)=
\frac{\partial F_A(x)}{\partial x_3}
\left(
\frac{\partial \t{y}_i}{\partial x_1}
\frac{\partial \t{y}_j}{\partial x_2}
-\frac{\partial \t{y}_i}{\partial x_2}
\frac{\partial \t{y}_j}{\partial x_1}\right)\,.
\ee

From Theorem 2 the Poisson algebra $B$ is given by
$$
[y_i,y_j]_{P.B.}=\sum_{i,j,k=1}^3\epsilon_{ijk}
\frac{\partial F_B(y)}{\partial y_k}\,.
$$
Hence if the smooth map $\t{y}(x)=(\t{y}_1,\t{y}_2,\t{y}_3)(x)$ 
gives rise to the algebra $B$, then
\be\label{18}
[\t{y}_i,\t{y}_j]_{P.B.}(x)=\sum_{i,j,k=1}^3\epsilon_{ijk}
\frac{\partial F_B(\t{y})}{\partial \t{y}_k}\,.
\ee
From (\ref{17p}) and (\ref{18}) we obtain
\be\label{19}
\frac{\partial F_A(x)}{\partial x_3}
\left(\frac{\partial \t{y}_i}{\partial x_1}
\frac{\partial \t{y}_j}{\partial x_2}
-\frac{\partial \t{y}_i}{\partial x_2}
\frac{\partial \t{y}_j}{\partial x_1}\right)
=\sum_{i,j,k=1}^3\epsilon_{ijk}
\frac{\partial F_B(\t{y})}{\partial \t{y}_k}\,.
\ee

(\ref{19}) has three different equations for $i=1,\,j=2$ resp.
$i=2,\,j=3$ resp. $i=3,\,j=1$.
Multiplying these equations by 
$\frac{\partial \t{y}_3}{\partial x_l}$ resp.
$\frac{\partial \t{y}_1}{\partial x_l}$ resp.
$\frac{\partial \t{y}_2}{\partial x_l}$, $l=1,2$, and
summing these equations together we get
$$
\frac{\partial F_B(\t{y})}{\partial x_l}=0\,,~~~~~l=1,2.
$$
Therefore $F_B(\t{y})$ is independent of $x_l$, $l=1,2$, and 
$F_B(\t{y})=constant$. This constant can be taken to be zero, since 
addition of a constant does not change the Poisson algebraic
structure of the manifold. $\rule{2mm}{2mm}$

{\sf [Therom 4].} If the map $\t{y}(x)=(\t{y}_1,\t{y}_2,\t{y}_3)(x)$
satisfies $F_B(\t{y})=0$, where $x$ satisfies $F_A(x)=0$, 
then $\t{y}$ generates the Poisson algebra $B$.

{\sf [Proof].} From Theorem 2 we know that there is a unique Poisson
algebra $B$ associated with the manifold $M_B$
(up to the algebraic equivalence given in definition 2). 
Hence if $\t{y}$ satisfies $F_B(\t{y})=0$,
then $\t{y}$ generates the algebra $B$. $\rule{2mm}{2mm}$

Summarizing, we have discussed the relations between integrable
Poisson algebraic structures and two-dimensional manifolds and have proved
that there is a unique relation between integrable Poisson algebras and 
two-dimensional smooth manifolds. We have also shown that the sufficient and
necessary condition for a smooth Poisson 
algebraic map $\t{y}(x)$ to act from an integrable Poisson algebra $A$ into 
an integrable Poisson algebra $B$ is that both 
$F_A(x)=0$ and $F_B(\t{y})=0$ are satisfied. These conclusions can be 
extended to the infinite dimensional case, see \cite{afc}.

\vspace{2.5ex}
ACKNOWLEDGEMENTS: We would like to thank the A.v. Humboldt
Foundation for the financial support given to the second named author.


\vspace{2.5ex}
\begin{thebibliography}{20}

\bibitem{found}
R. Abraham and J.E. Marsden, {\it Foundations of Mechanics}, 2nd ed.
Addision-Wesley, Benjamin/Cummings, Reagings, Mass.

\bibitem{moyal} 
J.E. Moyal, {\it Quantum Mechanics as a Statistical Theory}, 
Proc. Cambridge Phil. Soc. 45 (1949) 99.

\bibitem{poisson}
M.V. Karasev and V.P. Maslov, {\it Nonlinear Poisson Brackets, Geometry and 
Quantization}, Translations of Mathematical Monographs, Vol. 119,
America Mathematical Society, 1993.

\bibitem{poisson1}
K.H. Bhaskara and K. Viswanath, {\it Poisson Algebras and Poisson
Manifolds}, Pitman Research Notes in Mathematics Series 174, 1988.

\bibitem{geom}
J. Sniatycki, {\it Geometric Quantization and Quantum 
     Mechanics}, Springer Verlag, 1980.

\bibitem{wood}
N. Woodhouse, {\it Geometric Quantization}, Oxford: Clarendon Press,
      1980.

\bibitem{hofer}
H. Hofer and E. Zehnder, {\it Symplectic Invariants and Hamiltonian
Dynamics}, Birkh\"auser Verlag 1994.

\bibitem{symp}
B. Aebischer, M. Borer, M. K\"alin, Ch. Leuenberger and H.M. Reimann,
{\it Symplectic Geometry}, Progress in Mathematics, Vol. 124,
Birkh\"auser 1994.

\bibitem{afc}
S. Albeverio and S.M. Fei, 
{\it Current Algebraic Structures over Manifolds:
Poisson Algebras, q-Deformations and Quantization}, SFB-preprint, 1995.

\end{thebibliography}
\end{document}